\documentclass[12pt]{amsart}
\usepackage{graphicx}\usepackage{pdfpages}\usepackage{palatino}
\usepackage{tikz}
\usetikzlibrary{shapes.geometric, arrows}

 \theoremstyle{definition}
 
 \theoremstyle{remark}

\begin{document}
\title
{Coherent preorder of quantum states }

\author{Zhaofang Bai}\thanks{$^*$Corresponding author}
\email{baizhaofang@xmu.edu.cn} \address{School of Mathematical
Sciences, Xiamen University, Xiamen, Fujian, 361000, China}

\author{Shuanping Du$^*$}
\email{dushuanping@xmu.edu.cn} \address{School of Mathematical
Sciences, Xiamen University, Xiamen, Fujian, 361000, China}

\begin{abstract}

As an important quantum resource, quantum coherence play key role in quantum information processing. It is often concerned with  manipulation of families of quantum states rather than individual states in isolation.
Given two pairs of
coherent states $(\rho_1,\rho_2)$ and $(\sigma_1,\sigma_2)$, we are aimed to study how can we determine if there exists a strictly incoherent operation $\Phi$ such that $\Phi(\rho_i) =\sigma_i,i = 1,2$. This is also a classic question in quantum hypothesis testing.
In this note, structural characterization of coherent preorder under strongly incoherent operations is provided. Basing on the characterization, we propose an approach to realize coherence distillation  from rank-two mixed coherent states to $q$-level maximally coherent states.  In addition, one scheme of coherence manipulation between rank-two mixed states is also presented.

\end{abstract}

\maketitle

\section{Introduction }Quantum coherence is an essential physical resource which can be used to implement various tasks such as quantum computing
\cite{Hill}, cryptography \cite{Cole}, information processing
\cite{Stre1, Stre2, Diaz}, thermodynamics \cite{Lost}, metrology \cite{Fro}, and quantum biology \cite{Stre3}. Various efforts have been made to build the resource theory of coherence \cite{Abe}-\cite{Bau}. The resource theory of coherence consists of two fundamental elements:
free states and free operations. Free states are quantum states which can be prepared at no additional costs, while free operations
catch those physical transformations which can be carried out without consumption of resources.
Having confirmed the two main features, people initiate  investigation  of the corresponding theory, such as coherence manipulation and coherence quantification. One of the main advantages that
a resource theory offers is the lucid quantitative and operational
description as well as the manipulation of the relevant
resources at one's disposal.

Let us begin by recalling the
basic formalism of the resource theory of quantum coherence.  Free states are identified as incoherent states $$\rho=\sum_{i=1}^d\lambda_i|i\rangle\langle i|,$$ i.e., states which are diagonal in a fixed basis $\{|i\rangle\}_{i=1}^d$
for a $d$-dimensional system ${\mathcal H}$.
The set of incoherent states will be
labelled by ${\mathcal I}$. The choice of this basis depends on the particular problem under study, and in many relevant scenarios such a basis is naturally singled out by the unavoidable decoherence \cite{Zur}.

Free operations are identified as incoherent operations $(\text{ICO})$ which are specified by a finite set of $d\times d$ matrices $\{K_j\}$ satisfying $K_j\rho K_j^{\dag}/Tr(K_j\rho K_j^{\dag})\in {\mathcal
I}$ for all $\rho\in {\mathcal I}$, $$\Phi(\rho)=\sum_{j}K_j \rho K_j^{\dag}.$$ An incoherent operation can be interpreted as a measurement which can not create coherence even if one applies postselection on the measurement outcomes \cite{Bau}, we call such Kraus operators $\{K_j\}$
incoherent. Recall that, if both Kraus operators $K_j$ and $K_j^{\dag}$ are incoherent,
we call the operation strictly incoherent $(\text{SIO})$ \cite{Win, Gour2, Gu, Lud}. Different
definitions of free operations stemming from meaningful
physical considerations have been studied, such as  maximally incoherent operations
$(\text{MIO})$ \cite{Abe}, physically incoherent operations
$(\text{PIO})$ \cite{Gour3}, dephasing covariant incoherent operations $(\text{DIO})$ \cite{Gour3,Mar1},   genuinely incoherent operations $(\text{GIO})$ \cite{Vice}.  In spite of the
fact that the resource theory of coherence has found use in a
variety of practical settings \cite{Stre3}, there are no physically compelling free operations singled out, mirroring the
fundamental role of local operations and classical communication in the resource theory of entanglement \cite{Hor4}.

The class of strictly incoherent operations $(\text{SIO})$
appeared to be a promising candidate for a natural class of
operations satisfying desirable resource-theoretic criteria
while at the same time being motivated on physical grounds
and experimentally implementable, causing it to find
widespread use in the resource theory of coherence \cite{Win}-\cite{Was}.

The coherence manipulation is fundamental in the resource theory of quantum coherence.  It is aimed to study whether free operations introduce an order on the set of quantum states, i.e., whether, given two coherent states $\rho$ and $\sigma$,
either $\rho$ can be transformed into $\sigma$ or vice versa \cite{Stre3, Bau, Win, Gour2, Lud, Ben}, \cite{Du1}-\cite{FL1}. However, quantum
coherence theory is often concerned with the manipulation of families of quantum states
rather than individual states in isolation \cite{Nie1}. For instance, one needs to manipulate coherent states ${\rho}_1$ and ${\sigma}_1$ while freeze the other two coherent states ${\rho}_2$ and ${\sigma}_2$ in frozen quantum coherence \cite{BCA1}.
 The goal of the paper is how can we determine if there exists a SIO $\Phi$ such that
$\Phi(\rho_ i)=\sigma_i, i= 1,2$ for two pairs of coherent states $(\rho_1 ,\rho_2 )$ and $(\sigma_1 ,\sigma_2 )$.
A relevant question in quantum
hypothesis testing is how can we determine if there exists a quantum operation $\Phi$ such that
$\Phi(\rho_ i)=\sigma_i, i= 1,2.$
It also called  preorder of $(\rho_1 ,\rho_2 )$ and $(\sigma_1 ,\sigma_2 )$. This question was answered already in 1953 by Blackwell for the classical case \cite{Black}, and
in 1980 by Alberti and Uhlmann for the qubit case \cite{Albe1}. More recently, it was solved for pure states \cite{Chef1} and finally, it was fully solved \cite{Konig}-\cite{Bus4}. In \cite{Gour5}, it was shown that the preorder can be classified in terms of the conditional min-entropy.
Gour extended the definition of conditional min-entropy from bipartite quantum states to bipartite quantum operations which can be used to describe preorder of quantum operations \cite{Gour6}.

We will study the preorder question of pure states in coherence setting.
For two pairs of pure coherent states $(|\phi\rangle, |\psi\rangle)$ and $(|\alpha\rangle, |\beta\rangle)$, we are aimed to characterize when there exists a SIO $\Phi$ such that $\Phi(|\phi\rangle\langle\phi|)=|\alpha\rangle\langle\alpha|, \Phi(|\psi\rangle\langle\psi|)=|\beta\rangle\langle\beta|$.
Basing on the characterization, we present the coherence distillation scheme from rank-two mixed coherent states to $q$-level maximally coherent states under the strategy \cite{GLAY}.
This is a considerable progress in the study of coherence distillation of general mixed states \cite{Stre3, LZhou1, BFWG, Zao2}. Furthermore,
we also propose an approach to realize coherence manipulation between rank-two mixed states. It is known coherence manipulation between mixed states is a hard open question \cite{Gour2, Ben,Du2,LZhou1}.


\section{Preliminary}
Before stating our results, we need do some preparatory work. Firstly,  we recall the concept of coherent rank of pure state. The coherence rank of a pure state $|\phi\rangle$, denoted by $r(\phi)$, which is the number of basis elements for which $\phi_i\neq 0$ \cite{Ksp}. In analogy to the Schmidt rank in entanglement theory \cite{Hor4},
the coherent rank provides useful information about
the coherence content of a state and constrains the possible transformations among resource states. For instance, coherence rank can not increase under ICO and SIO \cite{Win, Gour2, Du1}. 
Secondly, for pure state $|\phi\rangle=\sum_{i=1}^d \phi_i|i\rangle$, the map which completely dephases in the
incoherent basis will be denoted by $\Delta$, and its action is
given by $$|\phi\rangle\mapsto \Delta(|\phi\rangle)=(|\phi_1|^2,|\phi_2|^2,\cdots, |\phi_d|^2)^t.$$
Finally, we collect some useful facts \cite{Bha}:

(i) For two real d-dimensional vectors $x=(x_1,x_2,\cdots, x_d)^t$ and
$y=(y_1,y_2,\cdots ,y_d)^t$,
 $x$ is majorized by $y$, written $x\prec y$, if for each
$k$ in the range $1, \cdots, d$, $\sum_i^k x_i^\downarrow\leq
\sum_i^k y_i^\downarrow$ with equality holding when $k= d$, and
where the $x_i^\downarrow$ indicates that elements are to be taken
in descending order. The majorization relation
is a partial order on pure states.

(ii) The majorization is well visualized by using the Lorenz curve. For vectors $x,y$, $x\prec y$ if and only if $x=D y$ for some
doubly stochastic matrix. Recall that a $d\times d$ matrix $D = (
d_{ij} )$ is called doubly stochastic if $d_{ij}\geq 0$ and
$\sum_{i=1}^d d_{ij}=\sum_{j=1}^d d_{ij}=1$.

(iii) Let $\pi$ be a permutation of $\{1,2,\cdots,d\}$. A $d\times d$
 matrix $P_\pi$ is the permutation matrix corresponding to
$\pi$ if it is obtained by permuting the rows  of the $d\times d$
identity matrix according to $\pi$. A permutation matrix has
exactly one entry 1 in each row and each column and 0 elsewhere.
It is known that Permutation matrices are unitary.

(iv) For every  doubly stochastic matrix $D$, it
can be written as a convex combination of permutation matrices, that is, there exist permutation matrices $P_{\pi_n}$ and
 probability coefficients $\lambda_n$ such that $D=\sum_n \lambda_nP_{\pi_n}$.\vspace{0.1in}

\section{coherent preorder}

We will characterize coherent preorder of $(|\phi\rangle, |\psi\rangle)$ and $(|\alpha\rangle, |\beta\rangle)$ in this section.
Specially, if $|\phi\rangle=|\alpha\rangle$, i.e., $|\phi\rangle$ is frozen under some SIO \cite{BCA1,Sompo,Fro2},
since majorization \cite{Gour2,Du1} rules the partial order of single-shot pure states, then a natural conjecture is that coherent preorder in this case can be described by d-majorization which is raised in various contexts including
mathematical statistics (comparison of statistical experiments)\cite{Blackwell1, Blackwell2,Cam1}, networks in
market \cite{Vein1}, chemical thermodynamics \cite{Ruch1, Ruch2},  mathematical and physical interests \cite{Albe1, Joe1, Marshall1}. Recall that for two pairs of probability distributions $(\Delta(|\phi\rangle), \Delta(|\psi\rangle))$ and $(\Delta(|\phi\rangle), \Delta(|\beta\rangle))$,
$(\Delta(|\phi\rangle), \Delta(|\beta\rangle))$ d-majorizes $(\Delta(|\phi\rangle), \Delta(|\psi\rangle))$ if and only if  there
exists a column-stochastic matrix $D$ with $D\Delta(|\phi\rangle)=\Delta(|\phi\rangle)$ and $D\Delta(|\beta\rangle)=\Delta(|\psi\rangle)$ \cite{Marshall1}.

The following theorem reveals the structure of coherent order in terms of relations of doubly stochastic matrices and superposed coefficients of pure states. This shows the structure of coherent order in general case is more sophisticated than d-majorization.

{\bf Theorem 1.} {\it Assume $r(\phi)=r(\alpha)$, $(|\phi\rangle,|\psi\rangle)\xrightarrow{{\rm
SIO}}(|\alpha\rangle,|\beta\rangle)$ if and only if there are doubly stochastic matrices $D_1,D_2$ and $c\in[0,1]$ satisfying the following conditions:\\
(i) $D_1\Delta(|\alpha\rangle)=\Delta(|\phi\rangle)$, $D_2\Delta(|\beta\rangle)=\Delta(|\psi\rangle)$, \\
(ii) $D_2=c^2D_1+(1-c^2)T$, for some doubly stochastic matrix $T$. And there are permutations $\pi_1$, $\pi_2$ such that \\
$P_{\pi_1}D_1P_{\pi_2}=\left(\begin{array}{ccccc}
   D_{11} & D_{12}  & 0& 0& 0\\
  D_{21}& D_{22} & 0& 0& 0\\
  0& 0 & D_{33} & D_{34} & 0\\
  0& 0 & D_{43} & D_{44} & 0\\
  0& 0 & 0&0 & D_{55} \end{array}
\right)$, $P_{\pi_1}TP_{\pi_2}=\left(\begin{array}{ccccc}
   0 & 0 & T_{13}& T_{14} & 0\\
  0& T_{22} &0& 0& 0\\
  0& 0 & T_{33}& T_{34}& 0\\
  T_{41}& 0 & T_{43}& T_{44}& 0\\
  0& 0 & 0&0 & T_{55}\end{array}
\right)$ \\ according to the space decomposition $H=\oplus_{i=1}^5  H_i$,
\begin{center}$\begin{array}{l}
H_1=\text{ span}\{|i\rangle\ \mid \phi_i\neq 0, \psi_i\neq 0\}\\
H_2=\text{ span}\{|i\rangle\ \mid \phi_i\neq 0,\psi_i=0\}\\
H_3=\text{ span}\{|i\rangle\ \mid \phi_i=0,\psi_i\neq 0, \beta_i\neq 0\}\\
H_4=\text{ span}\{|i\rangle\ \mid \phi_i=0,\psi_i\neq 0, \beta_i= 0\}\\
H_5=\text{ span}\{|i\rangle\ \mid \phi_i=0,\psi_i= 0, \beta_i= 0\}
\end{array}$\end{center}
(iii) $\frac{\beta_{\pi_n^{-1}(i)}}{\psi_i}=t\frac{\alpha_{\pi_n^{-1}(i)}}{\phi_i}$, $\phi_i\neq 0, \psi_i\neq 0$, here $|t|\leq1$ and  $\left(\begin{array}{ccccc}
   D_{11}  & D_{12}  \\
  D_{21} & D_{22} \end{array}\right)=\sum_n \lambda_n P_{\pi_n}$.}

\vspace{0.1in}

{\bf Remark 1.} 
If the energy of states are defined and both $\Delta(|\phi\rangle)$ and $\Delta(|\alpha\rangle)$ are the Gibbs distribution, i.e.,
$\Delta(|\phi\rangle)=\Delta(|\alpha\rangle)={p}^{Gibbs}$, then condition (i) is  thermo-majorization  between $(\Delta(|\phi\rangle),\Delta(|\psi\rangle))$ and $(\Delta(|\phi\rangle), \Delta(|\beta\rangle))$ (from the proof of Theorem 1). Thermo-majorization and other related problems in majorization have been intensively studied as
a possible extension of thermodynamics to small systems \cite{Hor3}-\cite{Shira}, which now becomes a major research field in quantum information theory known as quantum thermodynamics.
Condition (ii) reveals the relationship of doubly stochastic matrices which are induced by coherence manipulation of single-shot pure states.
It reflects essential difference between coherence order of two pairs of pure states and partial order of single-shot pure states.\vspace{0.1in}

For sufficiency of Theorem 1,  if $(\Delta(|\phi\rangle),\Delta(|\psi\rangle))$ and $(\Delta(|\alpha\rangle), \Delta(|\beta\rangle))$ share a common doubly stochastic
matrice, then the condition $r(\phi)=r(\alpha)$ is redundant as the following theorem shows.\vspace{0.1in}

{\bf Theorem 2.} {\it If there is a doubly stochastic matrix $D$ such that $D\Delta(|\alpha\rangle)=\Delta(|\phi\rangle)$, $D\Delta(|\beta\rangle)=\Delta(|\psi\rangle)$ and $\frac{\beta_{\pi_n^{-1}(i)}}{\psi_i}=t\frac{\alpha_{\pi_n^{-1}(i)}}{\phi_i}$, $D=\sum_n \lambda_n P_{\pi_n}$.
then $(|\phi\rangle,|\psi\rangle)\xrightarrow{{\rm
SIO}}(|\alpha\rangle,|\beta\rangle).$}

\section{coherence distillation}

In the section, we apply Theorem 1 to coherence distillation. One of the central problems in the resource theory of
coherence is the coherence distillation \cite{Stre3, Bau, Win, Lud, BFWG}, \cite{YZC}-\cite{RLS} which is the process that extracts pure coherent states from
general states via free operations. Especially, in \cite{GLAY}, G. Torun etc. have
performed a strictly incoherent operation (SIO) on a pure state and
obtain any of all $q$-level ($q = 2, 3,\ldots, d$) maximally coherent states
$|\Psi_q\rangle =\sum_{i=1}^q \frac {1} {\sqrt{q}} |i\rangle $, or an incoherent state ($q = 1$). Indeed, for a pure state $|\Psi\rangle $, the authors in \cite{GLAY} have constructed an explicit SIO which transforms  $|\Psi\rangle $ into $\sum_q p_q|\Psi_q\rangle \langle \Psi_q|$ for some probability coefficients $p_q$.

Although many interesting results in coherence distillation have been obtained, there
are still some open fundamental questions remaining to be
solved. The coherence distillation of general mixed states has
been left as an open question. Now,  basing on Theorem 1, we  propose the distillation procedure from rank-two mixed coherent states to q-level maximally coherent states as the following steps (See Fig.1).

\tikzset{
  rect1/.style = {
    shape = rectangle,
    draw = green,
    text width = 3cm,
    align = center,
    minimum height = 1cm,
  }
}

\tikzstyle{startstop} = [rectangle, rounded corners, minimum width = 2cm, minimum height=1cm,text centered, draw = black]
\tikzstyle{io} = [trapezium, trapezium left angle=70, trapezium right angle=110, minimum width=2cm, minimum height=1cm, text centered, draw=black]
\tikzstyle{arrow} = [->,>=stealth] 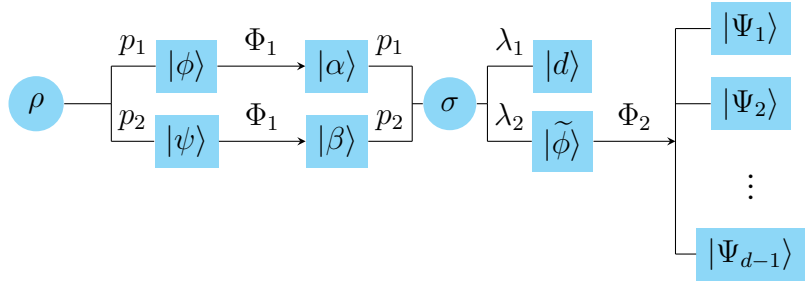
\begin{figure}
\begin{center}

\begin{tikzpicture}[node distance=1cm]

\node[circle, fill = cyan!40!white, ](1) at (0,0) {$\rho$};
\coordinate (c1) at (1,0);\coordinate (c2) at (1,0.5);\coordinate (c3) at (1,-0.5);
\draw [-] (1)--(c1);\draw [-] (c1)--(c2);\draw [-] (c1)--(c3);

\node[ rectangle, fill = cyan!40!white,   ] (2) at (2,0.5) {$|\phi\rangle$};
\node[rectangle, fill = cyan!40!white,  ](3) at (2,-0.5) {$|\psi\rangle$};
\node[rectangle, fill = cyan!40!white,  ](4) at (4,0.5){$|\alpha\rangle$};
\node[rectangle, fill = cyan!40!white,  ](5) at (4,-0.5){$|\beta\rangle$};
\coordinate (d1) at (5,0);\coordinate (d2) at (5,0.5);\coordinate (d3) at (5,-0.5);
\draw [-] (d2)--(d3);
\node[circle, fill = cyan!40!white, ](6) at (5.5,0) {$\sigma$}; \draw [-] (d1)--(6);

\coordinate (e1) at (6,0);\coordinate (e2) at (6,0.5);\coordinate (e3) at (6,-0.5);
\draw [-] (6)--(e1);\draw [-] (e1)--(e2);\draw [-] (e1)--(e3);
\node[rectangle, fill = cyan!40!white, ](7)at (7,0.5){$|d\rangle$};
\node[rectangle, fill = cyan!40!white, ](8)at (7,-0.5){$|\widetilde{\phi}\rangle$};

\coordinate (f1) at (8.5,0);\coordinate (f2) at (8.5,1);\coordinate (f3) at (8.5,-1);\coordinate (f4) at (8.5,-2);\coordinate (f5) at (8.5,-0.5);

\node[rectangle, fill = cyan!40!white,  ](9) at (9.5,1){$|\Psi_1\rangle$};
\node[rectangle, fill = white!40!white,  ](10)  at (9.5,-1) {$\vdots$};
\node[rectangle, fill = cyan!40!white,  ](11)  at (9.5,-2){$|\Psi_{d-1}\rangle$};
\node[rectangle, fill = cyan!40!white, ](12) at (9.5,0){$|\Psi_2\rangle$};

\draw[-](c2) -- node[above]{$p_1$}(2);
\draw[-](c3) -- node[above]{$p_2$}(3);
\draw[arrow](2) -- node[above]{$\Phi_1$}(4);
\draw[arrow](3) -- node[above]{$\Phi_1$}(5);
\draw[-](d2) -- node[above]{$p_1$}(4);
\draw[-](d3) -- node[above]{$p_2$}(5);
\draw[-](e2) -- node[above]{$\lambda_1$}(7);
\draw[-](e3) -- node[above]{$\lambda_2$}(8);
\draw[arrow](8) -- node[above]{$\Phi_2$}(f5);
\draw[-](f2) -- (9);\draw[-](f1) -- (12);\draw[-](f4) -- (11);\draw[-](f2) -- (f4);
\end{tikzpicture}
\end{center}
\caption{\small{For a rank-two mixed state $\rho=p_1|\phi\rangle\langle\phi|+p_2|\psi\rangle\langle\psi|$, we perform a SIO $\Phi_1$ sending $|\phi\rangle,|\psi\rangle$ to $|\alpha\rangle,|\beta\rangle$. $\sigma=p_1|\alpha\rangle\langle\alpha|+p_2|\beta\rangle\langle\beta|=\lambda_1 |d\rangle\langle d|+\lambda_2 |\widetilde{\phi}\rangle\langle\widetilde{\phi}|$, here $|\widetilde{\phi}\rangle=\sum_{i=1}^{d-1} \widetilde{\phi_i} |i\rangle$. Then one can apply the strategy in \cite{GLAY} to $|\widetilde{\phi}\rangle$ and obtain any of all $q$-level ($q = 2, 3, \ldots , d$) maximally coherent states $\Psi_q$, or an incoherent state (q = 1).}}

\end{figure}

{\bf Theorem 3.} {\it For a $2d-$dimentional state $\rho=p_1|\phi\rangle\langle\phi|+p_2|\psi\rangle\langle\psi|$ with
$$\sqrt{p_1}|\phi\rangle=\left(\begin{array}{c} \sqrt{\lambda _1}\sin \gamma \phi_1\\ \sqrt{\lambda _1}\sin \gamma \phi_2\\ \vdots\\ \sqrt{\lambda _1}\sin \gamma \phi_{d-1}\\ \sqrt{\lambda_2}\cos \gamma \\ 0 \\ \vdots\\ 0\end{array}
\right), \quad
\sqrt{p_2}|\psi\rangle=\left(\begin{array}{c} c\sqrt{\lambda _1}\cos \gamma \phi_1\\ c\sqrt{\lambda _1}\cos \gamma \phi_2\\ \vdots\\ c\sqrt{\lambda _1}\cos \gamma \phi_{d-1}\\ -c\sqrt{\lambda_2}\sin \gamma \\ \sqrt{1-c^2}\sqrt{\lambda _1}\cos \gamma \psi_1\\ \sqrt{1-c^2}\sqrt{\lambda _1}\cos \gamma \psi_2\\ \vdots\\ \sqrt{1-c^2}\sqrt{\lambda _1}\cos \gamma \psi_{d-1}\\ -\sqrt{1-c^2}\sqrt{\lambda_2}\sin \gamma \end{array}
\right),$$
here $\gamma \in(0,\frac {\pi} 4)$, $\sum_{i=1}^{d-1}|\phi_i |^2=\sum_{i=1}^{d-1}|\psi_i |^2=1$, $\phi_i\psi_i\neq 0$, $\lambda_1+\lambda_2=1, 0<c<1$,  there is a SIO $\Phi$ such that $\Phi(\rho)=\sum_{q=1}^{d-1} p_q|\Psi_q\rangle\langle \Psi_q|$ for some probability coefficients $p_q$. That is, we perform the distillation procedure from $\rho$ to q-level maximally coherent states.}

\section{coherence manipulation}

The study of coherence manipulation is moving ahead since the question is proposed \cite{Bau}.
A. Streltsov etc. \cite{Ben} have obtained that,  for  qubit states $\rho,\sigma$ with Bloch vector $r = (r_x,r_y,r_z)^T $ and $s = (s_x,s_y,s_z)^T$, $\rho$ can be converted into $\sigma$ by ICO if and only if $$ s^2 _x + s^2 _y \leq r^2 _x + r^2_y,\qquad s^2 _z \leq 1-\frac{ 1-r^2_z}{ r^2_x + r^2_y} (s_x^2 + s_y^2).$$
In \cite[Theorem 1]{Du1}, we have answered the question
in terms of majorization for pure states :  For any unit vectors
$|\phi\rangle=\sum_{i=1}^d \phi_i|i\rangle,
|\psi\rangle=\sum_{i=1}^d \psi_i|i\rangle$,
 $$|\psi\rangle \xrightarrow{\text{ICO}}|\phi\rangle
\quad \text{iff} \quad \Delta(|\psi\rangle)=(|\psi_1|^2,\cdots,
|\psi_d|^2)^t\prec\Delta(|\phi\rangle)=(|\phi_1|^2,\cdots, |\phi_d|^2)^t.$$
  Recently, Liu etc. \cite{LZhou1} give a necessary and sufficient condition for a mixed state $\rho$ to be transformed into a pure coherent state $|\phi\rangle$ via SIO:  $$\rho \xrightarrow{\text{SIO}}|\phi\rangle
\quad \text{iff} \quad \Delta(|\psi\rangle_\alpha)\prec\Delta(|\phi\rangle),$$
 here $|\psi\rangle_\alpha=\frac{P_\alpha \rho P_\alpha}{tr(P_\alpha \rho P_\alpha)}$, $\{P_\alpha\}$ is an orthogonal and complete set of incoherent projectors. In \cite{FL1}, the authors derive the lower bounds on the error of converting any full-rank coherent state to any  pure coherent state with certain probability by any free operation. Given any full-rank coherent state $\rho$ and any pure target coherent state
 $|\phi\rangle$,
 $$p\{\rho\rightarrow |\phi\rangle, \epsilon\}\Rightarrow\frac{\epsilon}{p}\geq\frac{\lambda_{\min}(\rho)(1-f_{|\phi\rangle})}{1+R(\rho)},$$ where $\lambda_{\min}(\rho)$ is the smallest eigenvalue of $\rho$, $f_{|\phi\rangle}=\max_{\sigma\in {\mathcal I}}tr(\sigma|\phi\rangle\langle\phi|)$ is the maximum overlap between $|\phi\rangle$ and incoherent states ${\mathcal I}$, $$R(\rho)=\min\{s|\exists s\geq 0, \text{state} \ \sigma \ \text{satisfying}\
 \frac{\rho+s\sigma}{1+s}\in{\mathcal I}\}$$ is the generalized robustness of state $\rho$. We build the no-go theorem of coherence manipulation in \cite{Du2}, i.e., finite number of measure conditions are insufficient to characterize coherence manipulation between general mixed states.  Therefore coherence manipulation between mixed states is complicated since it involves infinite number of conditions.

 Basing on Theorem 1, we propose an approach to
realize coherence manipulation between some rank-two mixed states. Our strategy is if $(|\phi\rangle, |\psi\rangle)$ and $(|\alpha\rangle, |\beta\rangle)$ share a SIO $\Phi$, then $$p|\phi\rangle\langle\phi|+(1-p)|\psi\rangle\langle\psi|\rangle \xrightarrow{\Phi}p|\alpha\rangle\langle\alpha|+(1-p)|\beta\rangle\langle\beta|,$$ $p\in[0,1]$.


{\bf Theorem 4.} {\it For $r\times r$ doubly stochastic matrices $D_{11}$ and $D_{21}$ and pure states $|\widetilde{\phi}\rangle$, $|\widetilde{\psi}\rangle$,$|\widetilde{\alpha}\rangle$,$|\widetilde{\beta}\rangle$,$|\widetilde{\tau}\rangle$ with coherent rank $r$ satisfying the following conditions
$$\begin{array}{l} D_{11}\Delta(\widetilde{\alpha})=\Delta(\widetilde{\phi}),\\
D_{11}\Delta(\widetilde{\beta})=\Delta(\widetilde{\psi}),\\ \frac{\beta_{\pi_n^{-1}(i)}}{\psi_i}=t\frac{\alpha_{\pi_n^{-1}(i)}}{\phi_i}, \text{ here } |t|\leq 1,D_{11}=  \sum_n \lambda_n P_{\pi_n}\\
D_{21}\Delta(\widetilde{\beta})=\Delta(\widetilde{\tau}),\end{array}$$ we have $$\left(\begin{array}{cc}p_1|\widetilde{\phi}\rangle \langle \widetilde{\phi}|+p_2|\widetilde{\psi}\rangle \langle \widetilde{\psi}|& p_2|\widetilde{\psi}\rangle \langle \widetilde{\tau}|\\
p_2|\widetilde{\tau}\rangle \langle \widetilde{\phi}| & p_2|\widetilde{\tau}\rangle \langle \widetilde{\tau}| \end{array}\right)\xrightarrow{\rm SIO} \left(\begin{array}{cc}p_1|\widetilde{\alpha}\rangle \langle \widetilde{\alpha}|+p_2|\widetilde{\beta}\rangle \langle \widetilde{\beta}|& 0\\0&0\end{array}\right).$$}

Indeed in Theorem 4, $H_2=H_3=H_5=0$, and
$\dim H_1=\dim H_4=r=\frac d 2$. Let $D_1=\left(\begin{array}{cc} D_{11}  &0 \\ 0& I\end{array}\right)$, $D_2=\left(\begin{array}{cc} c^2D_{11}  &(1-c^2)I  \\ (1-c^2)D_{21}& c^2I\end{array}\right)$, here $I$ is the $r\times r$ identity matrix. By Theorem 1, one can obtain Theorem 4 directly.

\section {Conclusions}

For two pairs of
coherent states $(\rho_1,\rho_2)$ and $(\sigma_1,\sigma_2)$, we study coherent order of two pairs of coherent states under SIOs.
The structural characterization of coherent order between pure states  is provided.
On this basis, coherence distillation scheme from rank-two mixed coherent stares to $q$-level maximally coherent states are offered.
Furthermore, coherence manipulation between rank-two mixed states can also be realized.

\vspace{0.1in}
{\it Acknowledgement---}
The authors thank the referee for his/her valuable suggestions which improve the presentation of the manuscript.
We acknowledge that the research was  supported by NSF of China (11671332), NSF of
Fujian (2018J01006).

\vspace{0.1in}
{\it Appendix---}
Proof of main results.

{\bf Proof of Theorem 1.}
Let $r=r(\phi)=r(\alpha)$.
Note that $(|\phi\rangle, |\psi\rangle)\xrightarrow{\text{SIO}}(|\alpha\rangle, |\beta\rangle)\Leftrightarrow (U|\phi\rangle, U|\psi\rangle)\xrightarrow{\text{SIO}}(V|\alpha\rangle, V|\beta\rangle)$ for permutation matrices $U, V$.
Without loss of generality, we assume \\
$|\phi\rangle =\left(\begin{array}{c}
                    \phi_1\\ \phi_2\\ \vdots\\ \phi_r\\0\\ \vdots \\ 0 \end{array}
                    \right),$
$|\alpha\rangle =\left(\begin{array}{c}
                    \alpha_1\\ \alpha_2\\ \vdots\\ \alpha_r\\0\\ \vdots \\ 0 \end{array}
                    \right),$
$|\psi\rangle =\left(\begin{array}{c}
                    \psi_1\\ \psi_2\\ \vdots\\ \psi_{s_1}\\0 \\ \vdots \\ 0\\ \psi_{r+1}\\ \vdots \\\psi_{r+s_2}\\ 0\\ \vdots \\ 0\end{array}
                    \right),$
$|\beta\rangle =\left(\begin{array}{c}
                    \beta_1\\ \beta_2\\ \vdots\\ \beta_{t_1}\\0\\ \vdots \\0 \\ \beta_{r+1}\\ \vdots \\\beta_{r+t_2}\\ 0\\ \vdots \\ 0\end{array}
                    \right).$   \\

  ``$\Rightarrow$": Assume that there exists some
strictly incoherent operation $\Phi$ such that $(|\phi\rangle, |\psi\rangle)\xrightarrow{\Phi}(|\alpha\rangle, |\beta\rangle)$.
Suppose the considered SIO $\Phi$ has Kraus operators $K_n$. It is evident that $K_n|\phi\rangle=\gamma_n|\alpha\rangle, K_n|\psi\rangle=\delta_n|\beta\rangle$ for some scalars $\gamma_n,\delta_n$.
Let
            $$\delta_{i,\pi(j)}=\left\{\begin{array}{cc}
            1,&\pi(j)=i\\
            0,&\pi(j)\neq i\end{array}\right.,\ \ \pi \text{ is a permutation}.$$
By the definition of SIO, one can write $K_n=P_{\pi_n}^{\dag}\text {diag}(k_1^{(n)}, k_2^{(n)},\ldots,k_d^{(n)})$, $P_{\pi_n}=(\delta_{i,\pi_n(j)})$. It follows that
$$k_{\pi_n(i)}^{(n)}\phi_{\pi_n(i)}=\gamma_n \alpha_i,$$
$$k_{\pi_n(i)}^{(n)}\psi_{\pi_n(i)}=\delta_n \beta_i.$$ Furthermore $s_1=t_1$ and  $s_2\geq t_2$. Note that there are $|\mu^{(n)}\rangle=\sum_{i=1}^{r+s_2}\mu_i|i\rangle$, $|\nu^{(n)}\rangle=\sum_{i=1}^{r+t_2}\nu_i|i\rangle$ with $\mu_i\nu_i\neq 0 $ and $K_n |\mu^{(n)}\rangle=\tau_n |\nu^{(n)}\rangle$. Therefore we have $K_{5 1}^{(n)}=K_{5 2}^{(n)}=K_{5 3}^{(n)}=K_{5 4}^{(n)}=0$ for $K_n=(K_{ij}^{(n)})$
according to the space decomposition $H=\oplus_{i=1}^5  H_i$. In addition, from $\phi_i=\psi_i=0, i=r+s_2+1,\cdots d$, we can take $K_{15}^{(n)}=K_{25}^{(n)}=K_{35}^{(n)}=K_{45}^{(n)}=0$.

{\bf Case I.} $r(\phi)=r(\alpha)=d$.

Since every column and every arrow of $K_n$ is
with at most 1 nonzero entry, we can obtain that
$\alpha_i\neq 0$. $k_i^{(n)}\neq 0$ for each $n,i$. And so $\gamma_n\neq 0, \delta_n\neq 0$. By a simple computation, one can see $$\frac{\psi_{\pi_n(i)}}{\phi_{\pi_n(i)}}=\frac {\delta_n}{\gamma_n}\frac{\beta_i}{\alpha_i},$$
$$\Pi_i \frac{\psi_{\pi_n(i)}}{\phi_{\pi_n(i)}}=(\frac {\delta_n}{\gamma_n})^d\Pi_i \frac{\beta_i}{\alpha_i}$$
$(\frac {\delta_n}{\gamma_n})^d$ is independently on $n$. Note that $\sum_n|\gamma_n|^2=\sum_n|\delta_n|^2=1$, we have $|\delta_n|=|\gamma_n|$.
Define $d_{ij}=\sum_{n,\pi_n^{-1}(i)=j}|\delta_n |^2, 1\leq i,j\leq d$,
then the matrix $D=(d_{ij})$
 is a doubly stochastic
matrix. By a direct computation, $D\Delta(|\alpha\rangle)=\Delta(|\phi\rangle)$ and $D\Delta(|\beta\rangle)=\Delta(|\psi\rangle)$. Moreover, $D=\sum_n |\delta_n|^2P_{\pi_n}$. In this case, $D_1=D_2=D$ and $c=1$.

{\bf Case II.} $r(\phi)=r(\alpha)=r<d$.

For the $n$ with $\gamma_n\neq 0$, by the property of SIO, $K_n=\left(\begin{array}{cc}
                               M_n & 0\\
                               0   & N_n \end{array}\right)$ according to the space decomposition $(H_1\oplus H_2)\oplus (H_3\oplus H_4\oplus H_5).$
In addition, $$K_n=\left(\begin{array}{ccccc}
K_{11}^{(n)} &K_{12}^{(n)}&0 &0 &0\\
K_{21}^{(n)}& K_{22}^{(n)}& 0& 0& 0\\
0 &0 & K_{33}^{(n)} & K_{34}^{(n)} &0\\
0& 0& K_{43}^{(n)} & K_{44}^{(n)} &0\\
0  & 0& 0& 0& K_{55}^{(n)}\end{array}\right)$$ according to the space decomposition $H_1\oplus H_2\oplus H_3\oplus H_4\oplus H_5.$
From $$K_n=P_{\pi_n}^{\dag}\text {diag}(k_1^{(n)}, k_2^{(n)},\ldots,k_d^{(n)}),$$ we have
\begin{equation} \begin{array}{ll} \pi_n^{-1}: &\{1,\ldots, s_1\}\rightarrow \{1,\ldots, s_1\}\\
& \{1,\ldots, r\}\rightarrow \{1,\ldots, r\}\\
                           &\{r+1,\ldots, r+s_2\}\rightarrow \{r+1,\ldots, r+s_2\}\\
                           & \{r+s_2+1,\ldots, d\}\rightarrow \{r+s_2+1,\ldots, d\}.\end{array}\label{1}\end{equation}
 Moreover, for  $i=1,\ldots, s_1$, $k_{\pi_n(i)}^{(n)}\neq 0$. So $$\frac{\psi_{\pi_n(i)}}{\phi_{\pi_n(i)}}=\frac {\delta_n}{\gamma_n}\frac{\beta_i}{\alpha_i},$$
$$\Pi_i \frac{\psi_{\pi_n(i)}}{\phi_{\pi_n(i)}}=(\frac {\delta_n}{\gamma_n})^{s_1}\Pi_i \frac{\beta_i}{\alpha_i},$$
$(\frac {\delta_n}{\gamma_n})^{s_1}$ is independently on $n$. Thus $|\delta_n|=c|\gamma_n|$. Note that
$$\begin{array}{lll}
   \sum_n|\delta_n|^2 & =& \sum_{\gamma_n\neq 0}|\delta_n|^2+\sum_{\gamma_n= 0}|\delta_n|^2\\
                      & =&\sum_{\gamma_n\neq 0}c^2|\gamma_n|^2+\sum_{\gamma_n= 0}|\delta_n|^2\\
                      & =& 1.\end{array}$$
This implies that $\sum_{\gamma_n\neq 0}c^2|\gamma_n|^2=c^2\sum_n|\gamma_n|^2=c^2\leq 1$. So $0<c\leq 1$, as desired.

For the $n$ with $\gamma_n= 0$, $\delta_n\neq 0$ (otherwise, we may assume $r+s_2=d$ and so $K_n=0$). Let $K_n=(K_{ij}^{(n)})$ according to the space decomposition $H=\oplus_{i=1}^5  H_i$. From $\gamma_n= 0$, we have $K_{ij}^{(n)}=0$, $i,j=1,2$. It is easy to see that $K_{23}^{(n)}=K_{24}^{(n)}=K_{25}^{(n)}=0$.  Combining with $\psi_i\neq0 (i=r+1,\ldots,r+s_2)$ and  $\beta_i= 0 $ $(i=r+t_2+1,\ldots,r+s_2)$, it follows that $K_{43}^{(n)}=K_{44}^{(n)}=0$. That is
$$K_n=\left(\begin{array}{ccccc}
0 & 0& K_{13}^{(n)} & K_{14}^{(n)} &0\\
0 & 0& 0& 0& 0\\
0 &0 & K_{33}^{(n)} & K_{34}^{(n)} &0\\
0& 0& 0 & 0 &0\\
0  & 0& 0& 0& K_{55}^{(n)}\end{array}\right).$$
On the other hand, $K_n=P_{\pi_n}^{\dag}\text {diag}(k_1^{(n)}, k_2^{(n)},\ldots,k_d^{(n)})$. Thus we can choose $\pi_n$ such that
\begin{equation}\begin{array}{ll} \pi_n^{-1}: & \{1,\ldots, s_1\}\rightarrow \{r+1,\ldots, r+s_2\}\\
                                & \{s_1+1,\ldots, r\}\rightarrow \{s_1+1,\ldots, r\}\\
                           &\{r+1,\ldots, r+t_2\}\rightarrow \{r+1,\ldots, r+s_2\}\\
                           & \{r+s_2+1,\ldots, d\}\rightarrow \{r+s_2+1,\ldots, d\}\end{array}
                           \label{2}\end{equation}

Define
 $d_{ij} ^{(1)}=\sum_{n,\pi_n^{-1}(i)=j}|\gamma_n |^2, 1\leq i,j\leq d$. Because $\sum_{n}|\gamma_n |^2=1$,
  $D_1=(d_{ij}^{(1)} )$  is doubly stochastic. From Eq.(\ref{1}), $D_1$ has the form $$\left(\begin{array}{ccccc}
   D_{11}  & D_{12}  & 0& 0& 0\\
  D_{21} & D_{22}  & 0& 0& 0\\
  0& 0 & D_{33} & D_{34} & 0\\
  0& 0 & D_{43} & D_{44} & 0\\
  0& 0 & 0&0 & D_{55} \end{array}\right).$$
Similarly, we can define the doubly stochastic matrix $D_2=(d_{ij}^{(2)}), 1\leq i,j\leq d $, by
$$\begin{array}{ll}
&d_{ij}^{(2)}=\sum_{n,\pi_n^{-1}(i)=j}|\delta_n |^2\\ = &\sum_{n,\pi_n^{-1}(i)=j,\gamma_n\neq 0}|\delta_n |^2+\sum_{n,\pi_n^{-1}(i)=j,\gamma_n= 0}|\delta_n |^2\\
=& c^2d_{ij}^{(1)} +(1-c^2)t_{ij},\end{array}$$  $t_{ij}=\frac 1{1-c^2} \sum_{n,\pi_n^{-1}(i)=j,\gamma_n= 0}|\delta_n |^2.$
$$T=(t_{ij})=\left(\begin{array}{ccccc}
   0 & 0 & T_{13}& T_{14}& 0\\
  0& T_{22} &0&0& 0\\
  0& 0 & T_{33}& T_{34}& 0\\
  T_{41}& 0 & T_{43}& T_{44}& 0\\
  0& 0 & 0&0 & T_{55}\end{array}\right)$$
Now, one can check that $D_1,D_2$ are the desired.

``$\Leftarrow$": Let $D_1=\sum_{n=1}^{N_1} \lambda_n P_{\pi_n}$, $T=\sum_{n=N_1+1}^{N_2} \lambda_n P_{\pi_n}$.
For $n=1,\ldots,N_1$, one can see $\pi_n$ satisfies conditions of $(2)$.
Define matrices $K_{11}^{(n)}=(k_{ij}^{(n)})$ $(i,j=1,\ldots,r)$ by
$$k_{ij}^{(n)}=\sqrt{\lambda_n}\delta_{j,\pi_n(i)}\frac{\alpha_i}{\phi_j}, $$
and $K_{22}^{(n)}=(k_{ij}^{(n)})$ $(i,j=r+1,\ldots,d)$) by
$$k_{ij}^{(n)}=\left\{\begin{array}{ll}
\sqrt{\lambda_n}\delta_{j,\pi_n(i)}\frac{\beta_i}{\psi_j} & i=r+1,\ldots,r+s_2 \\
\sqrt{\lambda_n}\delta_{j,\pi_n(i)} & i=r+s_2+1, \ldots, d.\end{array}\right.$$
Let $$K_n=\left(\begin{array}{cc} K_{11}^{(n)}&0\\ 0&c K_{22}^{(n)}\end{array}\right), n=1,2,\ldots, N_1.$$

For $n=N_1 +1,\ldots,N_2$, $\pi_n$ satisfies conditions of $(3)$.
Let
$$k_{ij}^{(n)}=\left\{\begin{array}{cl}
\sqrt{\lambda_n}\delta_{j,\pi_n(i)}\frac{\beta_i}{\psi_j} &\quad  i=1,\ldots, s_1, r+1,\dots,r+s_2  \\
0 &\quad i=s_1+1,\ldots, r\\
\sqrt{\lambda_n}\delta_{j,\pi_n(i)} &\quad  i=r+s_2+1,\ldots, d,\end{array}\right.$$
$$K_n=\sqrt{1-c^2}(k_{ij}^{(n)}).$$   A direct computation shows that the SIO which is specified by $K_n ( n=1,2, \ldots, N_2)$ can fulfill desired manipulation.\hfill $\square$

{\bf Proof of Theorem 2.}
Define Kraus operators $K_n=(k_{ij}^{(n)})$ by
$$k_{ij}^{(n)}=\left\{\begin{array}{ll}
\sqrt{\lambda_n}\delta_{j,\pi_n(i)}\frac{\alpha_i}{\phi_j} &  \phi_j\neq 0\\
\sqrt{\lambda_n}\delta_{j,\pi_n(i)}\frac{\beta_i}{\psi_j} & \phi_j=0, \psi_j\neq 0\\
\sqrt{\lambda_n}\delta_{j,\pi_n(i)} & \phi_j=\psi_j= 0.\end{array}\right.$$

One can check that SIO $\Phi$ represented by $\{K_n\}$ converts $(|\phi\rangle,|\psi\rangle)$ into $(|\alpha\rangle,|\beta\rangle)$.

{\bf Proof of Theorem 3.}
For any probability  vector $(|\alpha_1|^2,\ldots,|\alpha _{d-1}|^2)^t$ majorizing $(|\phi_1|^2,\ldots,|\phi _{d-1}|^2)^t$ and $(|\psi_1|^2,\ldots,|\psi _{d-1}|^2)^t$, we define
$$\sqrt{p_1}|\alpha\rangle =\left(\begin{array}{c} \sqrt{\lambda _1}\sin \gamma \alpha_1\\ \sqrt{\lambda _1}\sin \gamma \alpha_2\\ \vdots\\ \sqrt{\lambda _1}\sin \gamma \alpha_{d-1}\\ \sqrt{\lambda_2}\cos \gamma \\ 0 \\ \vdots\\ 0\end{array}
\right) , \quad
\sqrt{p_2}|\beta>=\left(\begin{array}{c} \sqrt{\lambda _1}\cos \gamma \alpha_1\\ \sqrt{\lambda _1}\cos \gamma \alpha_2\\ \vdots\\ \sqrt{\lambda _1}\cos \gamma \alpha_{d-1}\\ -\sqrt{\lambda_2}\sin \gamma \\ 0\\ \vdots\\ 0 \end{array}
\right).$$
Let $\widetilde{D_1}, \widetilde{D_2}$ be the doubly stochastic matrices with
$$\widetilde{D_1}\left(\begin{array}{c}|\alpha_1|^2\\\vdots\\|\alpha_{d-1}|^2\end{array}\right)=
\left(\begin{array}{c}|\phi_1|^2\\\vdots\\|\phi_{d-1}|^2\end{array}\right)\text{\ \   and \ \ } \widetilde{D_2}\left(\begin{array}{c}|\alpha_1|^2\\\vdots\\|\alpha_{d-1}|^2\end{array}\right)=
\left(\begin{array}{c}|\psi_1|^2\\\vdots\\|\psi_{d-1}|^2\end{array}\right).$$
Now define
$$D_1=\left(\begin{array}{cc} \left(\begin{array}{cc} \widetilde{D_1} &0 \\ 0 &1\end{array}\right) & 0\\ 0 &I_d\end{array}\right)\text{  and  }D_2=\left(\begin{array}{cc} c^2\left(\begin{array}{cc} \widetilde{D_1} & 0\\ 0&1\end{array}\right) & (1-c^2)I_d \\ (1-c^2)\left(\begin{array}{cc} \widetilde{D_2} & 0\\ 0&1\end{array}\right) &c^2I_d\end{array}\right).$$
One can check that $D_1$ and $D_2$ satisfy the conditions in Theorem 1. So there exists a SIO $\Phi_1$ sending $|\phi\rangle,|\psi\rangle$ to $|\alpha\rangle,|\beta\rangle$ separately.
On the other hand, it is easy to see that $\sigma=p_1|\alpha\rangle\langle\alpha|+p_2|\beta\rangle\langle\beta|=\lambda_1 |d\rangle\langle d|+\lambda_2 |\widetilde{\phi}\rangle\langle\widetilde{\phi}|$ with $|\widetilde{\phi}\rangle=\sum_{i=1}^{d-1} \alpha_i |i\rangle$. Without loss of generality, one may assume $\{|\alpha_i|\}_{i=1}^{d-1}$ is positive and decending. Applying the strategy from \cite{GLAY}, we define
$$\begin{array}{l}
 p_q=q(|\alpha_q|^2-|\alpha_{q+1}|^2), \quad q=1,2,\ldots,d-2,\\  p_{d-1}=d|\alpha_{d-1}|^2,\\
K_q=\sqrt{p_q}(\frac 1{\sqrt q}\sum_{i=1}^{q}\frac{|i\rangle \langle i|}{\alpha_i}+|d\rangle \langle d |)\quad q=1,2,\ldots,d-1.\end{array}$$
By construction, we
have that
$$\begin{array}{l}
K_q|\widetilde{\phi}\rangle  =\sqrt{p_q}|\Psi_q\rangle ,\\
K_q|d\rangle  =\sqrt{p_q}|d\rangle ,\\
\sum K_q^{\dag}K_q=I_d,\ \
\sum p_q=1.\end{array}$$
In addition,  the SIO $\sum K_q\cdot K_q^{\dag}$ denoted by $\Phi_2$ sends $\sigma$ to $\lambda_1 |d\rangle\langle d|+\lambda_2 \sum_{q=1}^{d-1}p_q|\Psi_q\rangle \langle \Psi_q|$. Therefore the composition $\Phi_2\circ\Phi_1$ is the desired transformation.

\end{document}